\documentclass[11pt]{article}
\usepackage[utf8]{inputenc}
\usepackage[T1]{fontenc}
\usepackage{lmodern}
\usepackage[margin=1in]{geometry}
\usepackage{microtype}
\usepackage{amsmath,amssymb,amsthm}
\usepackage{booktabs,tabularx,array,longtable}
\usepackage{enumitem}
\usepackage{xcolor}
\usepackage{hyperref}
\usepackage{graphicx}
\usepackage{tikz}
\usepackage{listings}
\usepackage{caption}
\usepackage{url}
\usetikzlibrary{calc}
\hypersetup{colorlinks=true,linkcolor=blue,citecolor=blue,urlcolor=blue,pdftitle={AEX: Non-Intrusive Multi-Hop Attestation and Provenance for LLM APIs}}

\setlist[itemize]{leftmargin=1.5em,topsep=0.2em,itemsep=0.2em}
\setlist[enumerate]{leftmargin=1.6em,topsep=0.2em,itemsep=0.2em}

\newcommand{\AEX}{\textsc{AEX}}
\newcommand{\JCS}{\mathrm{JCS}}
\newcommand{\SHA}{\mathrm{SHA\mbox{-}256}}
\newcommand{\Req}{\texttt{AEX-REQ-V1}}
\newcommand{\Resp}{\texttt{AEX-RESP-V1}}
\newcommand{\Chunk}{\texttt{AEX-CHUNK-V1}}
\newcommand{\Stream}{\texttt{AEX-STREAM-V1}}
\newcommand{\Transform}{\texttt{AEX-TRANSFORM-V1}}
\newcommand{\OriginOutput}{\texttt{AEX-ORIGIN-OUTPUT-V1}}
\newcommand{\OutputTransform}{\texttt{AEX-OUTPUT-TRANSFORM-V1}}
\newcommand{\SigPrefix}{\texttt{AEX-ATTESTATION-V1}}

\title{\AEX: Non-Intrusive Multi-Hop Attestation and Provenance for LLM APIs}
\date{\today}
\author{
  Yongjie Guan \\ Zhejiang University of Technology
}

\begin{document}
\maketitle

\begin{abstract}
Hosted large language models are increasingly accessed through remote APIs, but the API boundary still offers little direct evidence that a returned output actually corresponds to the client-visible request. Recent audits of \emph{shadow APIs} show that unofficial or intermediary endpoints can diverge from claimed behavior, while existing approaches such as fingerprinting, model-equality testing, verifiable inference, and TEE attestation either remain inferential or answer different questions. We propose \AEX, a non-intrusive attestation extension for existing JSON-based LLM APIs. \AEX\ preserves request, response, tool-calling, streaming, and error semantics, and instead adds a signed top-level \texttt{attestation} object that binds a client-visible request projection to either a complete response object or a committed streaming output. To support realistic deployments, \AEX\ provides explicit request-binding modes, signed request-transform receipts for trusted intermediaries, and source-output / output-transform receipts for trusted output rewriting. For streaming, it separates checkpoint proofs for verified prefixes of an unmodified source stream from complete-output lineage for outputs that have been rewritten, buffered, aggregated, or re-packaged, preventing transformed outputs from being mistaken for source-stream prefixes. \AEX\ therefore makes a deliberately narrow claim: a trusted issuer attests to a specific request-output relation, or to a specific complete-output lineage, at the API boundary. We present the protocol design, threat model, verification state machine, security and privacy analysis, an OpenAI-compatible chat-completions profile, and a reference TypeScript prototype with local conformance tests and microbenchmarks.
\end{abstract}

\section{Introduction}
Large language models (LLMs) have become infrastructure consumed primarily through remote APIs rather than local model execution \cite{gao2025met,openai-chat}. That shift moves an important trust boundary outward: users and downstream systems therefore rely on a provider's API boundary not only for availability and latency, but also for provenance, integrity, and reproducibility.

Recent evidence suggests that this trust boundary is increasingly stressed. Zhang et al. present the first systematic audit of LLM \emph{shadow APIs} --- unofficial endpoints that claim compatibility with official model APIs --- and identify 17 such services appearing in 187 papers. Their study reports utility divergence up to 47.21\%, fingerprint-verification failures on 45.83\% of evaluated endpoints, and concludes by calling for ``lightweight official verification endpoints'' that researchers could query to confirm model identity independently \cite{zhang2026shadow}. The same paper also shows that fingerprinting and statistical testing are useful but incomplete: in some cases, identity-consistent serving still fails to preserve behavior, suggesting that audit-by-inference is not enough \cite{zhang2026shadow}.

A growing body of work seeks to verify LLM services in other ways. LLM fingerprinting aims to infer which model likely sits behind an endpoint \cite{pasquini2025llmmap}. Model Equality Testing (MET) detects distributional mismatches between an endpoint and a reference model \cite{gao2025met}. Verifiable inference schemes such as SVIP ask providers to expose model-internal evidence \cite{sun2025svip}. TEE-based systems and remote-attestation protocols seek stronger guarantees about the environment in which code runs \cite{akama2024rawebs,menetrey2022attestation,zhang2025attestllm}. These are all useful, but they answer different questions.

The gap addressed in this paper is simpler and more deployment-oriented: \emph{how can an existing LLM API attach direct, online, verifiable evidence that a particular client-visible request is bound to a particular response object or streaming output sequence?}

We argue that the right place to answer that question is the API protocol boundary itself. In practice, many LLM APIs are JSON-based and frequently streamed over Server-Sent Events (SSE). They also often pass through middleware layers that add defaults, rewrite prompts, inject policy instructions, redact outputs, buffer streams, or re-encode responses. A deployable attestation layer therefore needs to satisfy eight constraints at once:
\begin{enumerate}
    \item preserve existing request and response semantics;
    \item bind a client-visible request object without requiring byte-for-byte preservation of the transport representation;
    \item support explicit, client-selected request-binding strictness;
    \item bind entire streaming outputs, including order, count, and completeness;
    \item detect truncation; and
    \item distinguish between unauthorized request mutation and trusted, explicitly accepted request rewriting;
    \item represent trusted output rewriting, re-encoding, aggregation, or stream re-packaging explicitly rather than treating it as undifferentiated tampering; and
    \item distinguish source-stream prefix proofs from complete-output lineage proofs so that clients do not over-interpret partial evidence.
\end{enumerate}

This paper proposes \AEX\ (Attestation EXtension), a protocol extension that addresses those constraints by composing well-understood mechanisms --- canonical JSON, cryptographic hashing, digital signatures, issuer-discovered public keys, and hash chaining --- into an LLM-specific transaction attestation layer. \AEX\ deliberately does \emph{not} claim a new cryptographic primitive. Its contribution is protocol design: a concrete, interoperable way to add request-output attestation to existing LLM APIs without changing their body semantics.

\paragraph{Contributions.}
This paper makes five contributions.
\begin{itemize}
    \item It formulates the \emph{API-boundary attestation} problem for hosted LLM services and argues that recent shadow-API evidence makes this problem practically urgent \cite{zhang2026shadow}.
    \item It presents \AEX, a non-invasive attestation extension for existing JSON-based LLM APIs, including request commitments, complete-output commitments, explicit request-binding modes, issuer-signed top-level attestations, and output-mode-aware lineage metadata.
    \item It introduces three LLM-specific protocol mechanisms: a \emph{dual-anchor streaming hash chain}, an explicit \emph{request-transform chain} for trusted intermediaries, and a source-output / output-transform receipt model for trusted output rewriting.
    \item It provides a security, privacy, and deployment analysis and positions \AEX\ against neighboring approaches from HTTP message signing, provenance frameworks, TEE attestation, and LLM auditing.
    \item It reports a reference TypeScript prototype together with simulator-based local validation, automated conformance testing, and scenario-driven microbenchmarks over a gateway/proxy/provider testbed that exercises both checkpoint-based streaming proofs and complete-output-lineage flows.
\end{itemize}

\paragraph{Scope.}
\AEX\ is a provenance and integrity mechanism at the transaction boundary. It proves that a trusted issuer signed a binding between a request commitment and an output commitment, and when output-lineage metadata is present it additionally proves that the final complete output was derived from an explicitly attested source output through a signed transform chain. It does \emph{not} prove that an output is true, that no hidden prompt exists, that the serving path used a specific set of model weights, or that the provider behaved honestly beyond what is captured by the signed object.

\section{Problem Statement and Threat Model}
\subsection{System model}
We consider a client invoking an LLM API whose request and response bodies are JSON objects. Responses may be non-streaming or may arrive as a sequence of JSON objects, often carried in SSE framing \cite{whatwg-sse,openai-chat-stream}. Between the client and the model-serving backend, a request may pass through multiple components:
\begin{itemize}
    \item API gateways or reverse proxies,
    \item middleware that injects defaults or policy text,
    \item auditing or logging sidecars,
    \item trusted-execution environments (TEEs) or attestation-capable serving stacks, and
    \item untrusted or semi-trusted third-party intermediaries.
\end{itemize}

An \emph{issuer} is the entity that signs the final attestation. A \emph{verifier} is the party that receives the response together with the original request and checks the attestation. A verifier may be the client itself, an SDK, a gateway, or an auditor.

\subsection{Goals}
\AEX\ is designed to satisfy the following goals.
\begin{enumerate}
    \item \textbf{Non-invasiveness.} The original API body semantics, tool semantics, streaming semantics, and error semantics remain unchanged.
    \item \textbf{Client-visible request binding.} The verifier can reconstruct a stable request commitment from the request object the client actually sent.
    \item \textbf{Explicit compatibility control.} The client can select how strict request binding should be, to account for benign middleware changes.
    \item \textbf{Streaming integrity.} The verifier can detect chunk insertion, deletion, reordering, and truncation.
    \item \textbf{Trusted request rewriting.} Explicitly trusted intermediaries can rewrite requests without collapsing all differences into a generic request mismatch.
    \item \textbf{Trusted output rewriting.} Explicitly trusted intermediaries can rewrite, buffer, re-encode, aggregate, or re-package outputs through an auditable output-lineage chain.
    \item \textbf{Sound streaming proof boundaries.} Prefix proofs for a source stream must be distinguished from complete-output lineage for transformed outputs; the protocol must not let clients confuse one for the other.
    \item \textbf{Deployment realism.} The mechanism should compose with OpenAI-compatible APIs and with issuer-specific key-discovery infrastructure.
\end{enumerate}

\subsection{Non-goals}
\AEX\ does not attempt to solve all trust problems in hosted LLM use.
\begin{itemize}
    \item It does \emph{not} prove model truthfulness or factual correctness.
    \item It does \emph{not} prove absence of hidden prompts, hidden retrieval, or hidden policy systems.
    \item It does \emph{not} prove exact model weights or hardware configuration.
    \item It does \emph{not} prove that every hidden intermediate output in an accepted output-transform chain is visible or reconstructible to the client.
    \item It does \emph{not} prove that an accepted output transform is semantically correct, reasonable, or minimal.
    \item It does \emph{not} replace TEE remote attestation, model fingerprinting, or model-equality testing; it complements them.
\end{itemize}

\subsection{Adversary model}
We consider an active network or middleware adversary that can tamper with requests and responses, reorder or truncate streaming outputs, attach stale or unrelated attestations, or attempt to induce verification failures through key-discovery abuse. We also consider an untrusted intermediary that may silently rewrite the request or misrepresent a backend model.

\AEX\ can detect many such actions if they alter the attested request or output relation. However, if the issuer itself is malicious, or if the verifier trusts the wrong issuer, the attestation proves only that \emph{that issuer} signed the relation. This is an unavoidable trust-root limitation shared by other attestation systems \cite{menetrey2022attestation,akama2024rawebs}.

\section{Design Overview}
\subsection{Boundary and representation}
\AEX\ operates over JSON \emph{semantic objects}, not raw transport bytes. This choice follows the logic of JSON canonicalization: cryptographic operations need an invariant representation even when the ``wire'' form differs in whitespace, field ordering, or numeric spelling \cite{rfc8785}. Accordingly, \AEX\ uses the JSON Canonicalization Scheme (JCS) \cite{rfc8785} as its canonical byte representation and Ed25519 signatures as a mandatory signature algorithm \cite{rfc8032}. The design assumes I-JSON-like inputs with unique object keys and no NaN/Infinity values \cite{rfc7493}; objects that cannot be deterministically canonicalized are out of scope for successful attestation. Numeric spellings such as \texttt{1}, \texttt{1.0}, and \texttt{1e3} therefore matter only through the canonicalized value they denote, not through their original lexical representation.

The protocol attaches a top-level field named \texttt{attestation} to a JSON response object. For commitment computation, the top-level \texttt{attestation} field is removed; nested fields with the same name are left untouched. This makes \AEX\ detached from the business payload while allowing the payload to remain an ordinary JSON object. Domain separation is provided by literal UTF-8 tags such as \Req, \Resp, \Chunk, \Stream, \Transform, \OriginOutput, \OutputTransform, and \SigPrefix, each prepended to the relevant canonicalized object before hashing or signing.

\subsection{Requests, outputs, and profiles}
A protocol profile defines how \AEX\ is attached to a particular API. The motivating profile in this paper is an OpenAI-compatible chat-completions API, which accepts a JSON request containing a list of messages and returns either a single chat-completion object or a streamed sequence of chunk objects \cite{openai-chat}. Because many deployed APIs follow this interface or close variants of it, an OpenAI-compatible profile provides a realistic baseline.

\subsection{Core commitments}
\AEX\ introduces three kinds of cryptographic commitments.
\begin{itemize}
    \item \textbf{Request commitment.} A digest over a client-visible request projection.
    \item \textbf{Effective request commitment.} An optional digest over the final, publicly explainable request actually executed after trusted rewriting or deterministic normalization.
    \item \textbf{Output commitment.} A digest over a complete non-streaming response object or over a complete authenticated streaming chain.
\end{itemize}

In all cases, commitments are represented in JSON as strings of the form
\begin{center}
\texttt{sha256:<64 lowercase hex characters>}
\end{center}
but are interpreted internally as raw 32-byte digests. Output commitments are always interpreted together with an explicit \texttt{output\_mode} so that complete non-streaming objects and complete streaming chains are never conflated.

\subsection{Request binding as an explicit choice}
A distinctive feature of \AEX\ is that the client chooses the request-binding mode. The reason is practical: some deployments require end-to-end request immutability, whereas others legitimately add fields such as trace identifiers or routing hints at the gateway. \AEX\ therefore defines three request-binding modes:
\begin{enumerate}
    \item \texttt{full}: bind the entire request object after removing the top-level \texttt{attestation};
    \item \texttt{top\_level\_exclude}: bind the request after removing a named set of top-level fields;
    \item \texttt{top\_level\_include}: bind only a named set of top-level fields.
\end{enumerate}

The third mode has one subtle but important feature: it also binds which listed top-level fields are \emph{absent}. This protects against a class of compatibility/security failures in which a middleware layer silently injects a previously absent field.

\subsection{Streaming as a first-class object}
Streaming is not merely a transport optimization; it changes what must be verified. In a non-streaming response, integrity of the final JSON object is sufficient. In a streamed response, the verifier must be able to detect not only tampering with any chunk but also deletion, insertion, reordering, and premature termination. Prior stream-authentication work has long recognized that efficient authentication for streamed data needs different machinery than independent signatures per packet \cite{golle2001streamauth}. \AEX\ adopts that lesson at the LLM API layer using a chunk hash chain.

However, \AEX\ draws a sharp line between two different streaming claims. If the delivered stream is the source issuer's own untransformed stream, checkpoint attestations can certify verified prefixes of that source stream. If a trusted intermediary changes chunk content, count, order, framing boundaries, aggregation, or streaming/non-streaming representation, then the final output is no longer the source stream. In that case, \AEX\ stops offering prefix proofs and instead offers only complete-output lineage: the final terminal attestation can identify a source issuer's original complete output and a signed chain of trusted output transforms that lead to the final complete output delivered to the client.

\subsection{Trusted request transforms}
Existing message-signing systems already acknowledge that intermediaries may transform messages. HTTP Message Signatures, for example, are designed to survive only those transformations that the signer intentionally excludes from the signed component set \cite{rfc9421}. \AEX\ goes further for LLM APIs by making trusted request mutation explicit through signed \texttt{request\_transforms} receipts. A sequence of such receipts can attest that a request commitment was transformed into a later effective-request commitment by named, trusted intermediaries. This turns otherwise opaque prompt rewriting into a verifiable chain rather than an implicit exception.

\subsection{Trusted output transforms and source receipts}
Request-side provenance alone is insufficient once gateways or trusted middleware buffer outputs, redact spans, apply safety filtering, convert a source stream into one non-streaming object, or re-package one complete object as a stream. \AEX\ therefore makes trusted output mutation explicit too. An \texttt{origin\_output} receipt states that a source issuer signed an original complete output commitment for the same request context. A sequence of \texttt{output\_transforms} receipts can then map that original complete output commitment into the final complete output commitment signed by the terminal issuer.

Each output-lineage node carries an explicit \texttt{output\_mode}, allowing the lineage to cross between non-streaming and streaming forms without ambiguity. This detail matters because a commitment over a full JSON object and a commitment over a full chunk sequence are different objects even when they encode similar business content. Importantly, \texttt{output\_transforms} attest only relations between \emph{complete} outputs. They are not prefix-to-prefix streaming proofs.

\begin{figure}[t]
\centering
\includegraphics[width=0.9\linewidth]{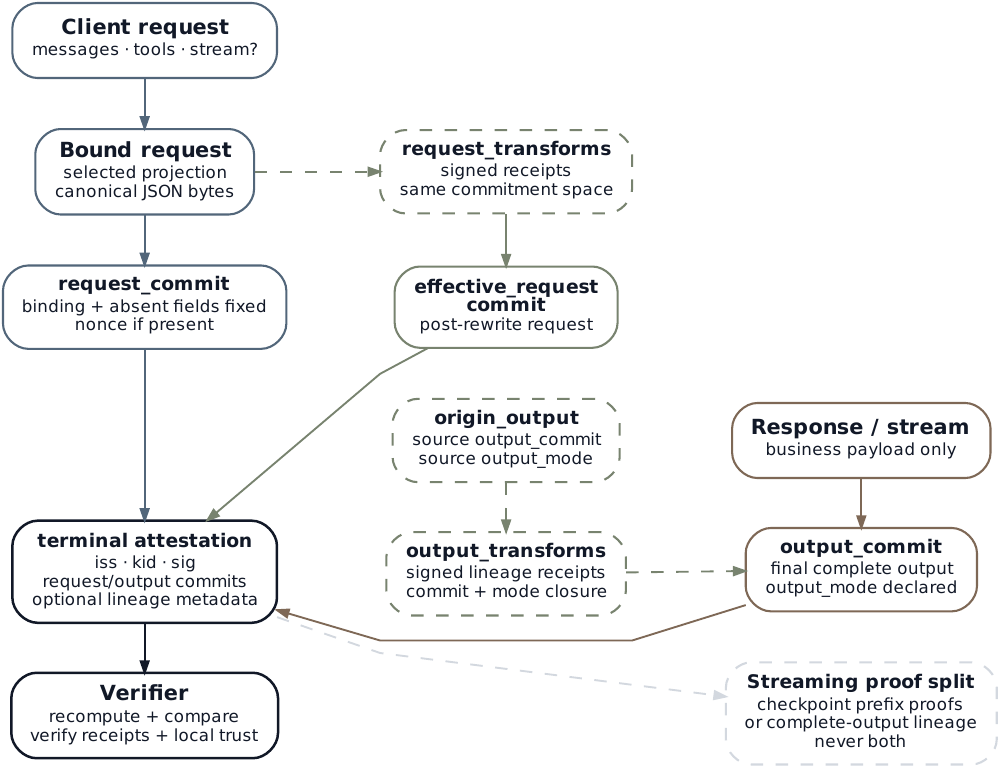}
\caption{High-level structure of \AEX. Request-side commitments flow from the client-visible request and optional signed \texttt{request\_transforms}; output-side lineage, when present, flows from a source output through \texttt{origin\_output} and \texttt{output\_transforms} to the final \texttt{output\_commit}. The terminal attestation binds these commitments to the returned response or stream without changing business-payload semantics.}
\label{fig:aex-overview}
\end{figure}

Figure~\ref{fig:aex-overview} summarizes the protocol boundary: \AEX\ leaves the business payload in its existing API shape while adding a detached terminal attestation that binds request-side commitments, and when present complete-output lineage metadata, to the returned output.

\section{Protocol Mechanics}
\subsection{Activation}
\AEX\ is enabled by adding a top-level \texttt{attestation} member to the request. The minimal enabling form is simply
\begin{lstlisting}
"attestation": true
\end{lstlisting}
A parameterized form may include a nonce, a required flag, and a request-binding object. The nonce is supplied by the client and echoed back by the issuer. It reduces linkability and can support replay detection when combined with verifier-side freshness or stateful replay policy. The \texttt{required} flag is a no-silent-downgrade control: if a client marks attestation as required, a service that cannot honor the requested \AEX\ behavior should fail explicitly rather than return an unattested success response. When a client expects complete-output lineage metadata such as \texttt{origin\_output} or \texttt{output\_transforms}, high-entropy nonces become especially valuable because they reduce correlatability across repeated requests.

\subsection{Canonical request input}
Let $R$ denote the original request object after removing the top-level \texttt{attestation}. Let $B$ denote the binding descriptor chosen by the client. Let $N$ denote an optional request-binding nonce. \AEX\ computes a canonical request input object
\[
\mathrm{BoundRequestInput}(R,B,N)
\]
that contains the binding descriptor, the selected request projection, and (if present) the nonce. The top-level request \texttt{attestation} object itself is never part of the request projection; it contributes only the extracted nonce when one is present.

For \texttt{top\_level\_include}, the object also contains an \texttt{absent\_fields} array listing requested top-level fields that were not present in $R$. This is the mechanism that binds the distinction between ``present'' and ``not present'' and prevents a middleware layer from silently injecting a previously absent protected field.

The request commitment is then defined as
\[
\mathrm{request\_commit} = \SHA\bigl(\Req \;\|\|\; \JCS(\mathrm{BoundRequestInput}(R,B,N))\bigr).
\]

\subsection{Effective request commitment}
Some services deterministically derive an effective request that differs from the externally visible request: they may fill in documented defaults, normalize deprecated parameter spellings, or apply trusted prompt rewriting. To represent such behavior without redefining the client-visible request, \AEX\ allows an optional \texttt{effective\_request\_commit}. This commitment is computed in the \emph{same} commitment space as the original request commitment:
\[
\mathrm{effective\_request\_commit} = \SHA\bigl(\Req \;\|\|\; \JCS(\mathrm{BoundRequestInput}(R_{\mathrm{eff}},B,N))\bigr).
\]

The same binding descriptor and the same nonce must be used for both. This matters because it makes the original request, the effective request, and every request-transform step comparable within one commitment space.

\subsection{Output modes and complete-output commitments}
\AEX\ distinguishes two complete-output modes:
\begin{itemize}
    \item \texttt{non\_stream}: the complete output is one top-level JSON object; and
    \item \texttt{stream}: the complete output is a full committed sequence of JSON chunk objects.
\end{itemize}
The \texttt{output\_mode} is therefore part of the meaning of every complete-output commitment.

For a non-streaming response object $O$, remove the top-level \texttt{attestation} and compute
\[
\mathrm{output\_commit} = \SHA\bigl(\Resp \;\|\|\; \JCS(O)\bigr).
\]

Streaming is defined over \emph{JSON event objects}, not arbitrary transport frames. If SSE is used, the relevant unit is a complete SSE event whose concatenated \texttt{data:} payload parses as a JSON object \cite{whatwg-sse}. Non-JSON markers such as \texttt{[DONE]} are outside the commitment chain. Let the $i$th committed chunk after removal of its top-level \texttt{attestation} be $C_i$. Define
\[
H_i = \SHA\bigl(\Chunk \;\|\|\; \mathrm{uint64be}(i) \;\|\|\; \JCS(C_i)\bigr).
\]
Here $i$ starts at 1 in arrival order, and \(\mathrm{uint64be}(i)\) denotes an 8-byte unsigned big-endian encoding.

Let $r$ be the raw 32-byte request commitment, and let $e$ be the raw 32-byte effective-request commitment if present, otherwise $e=r$. Initialize the chain as
\[
\mathrm{chain}_0 = \SHA\bigl(\Stream \;\|\|\; r \;\|\|\; e\bigr).
\]
Then update recursively:
\[
\mathrm{chain}_i = \SHA\bigl(\mathrm{chain}_{i-1} \;\|\|\; H_i\bigr).
\]
After $n$ committed chunks, the streaming complete-output commitment is
\[
\mathrm{output\_commit} = \mathrm{chain}_n.
\]

This construction makes the complete streaming commitment sensitive to chunk content, order, count, and the request anchors. Any insertion, deletion, mutation, reordering, re-chunking, aggregation, or split of JSON chunk objects changes the committed object and therefore constitutes an output transform.

\begin{figure}[t]
\centering
\includegraphics[width=0.9\linewidth]{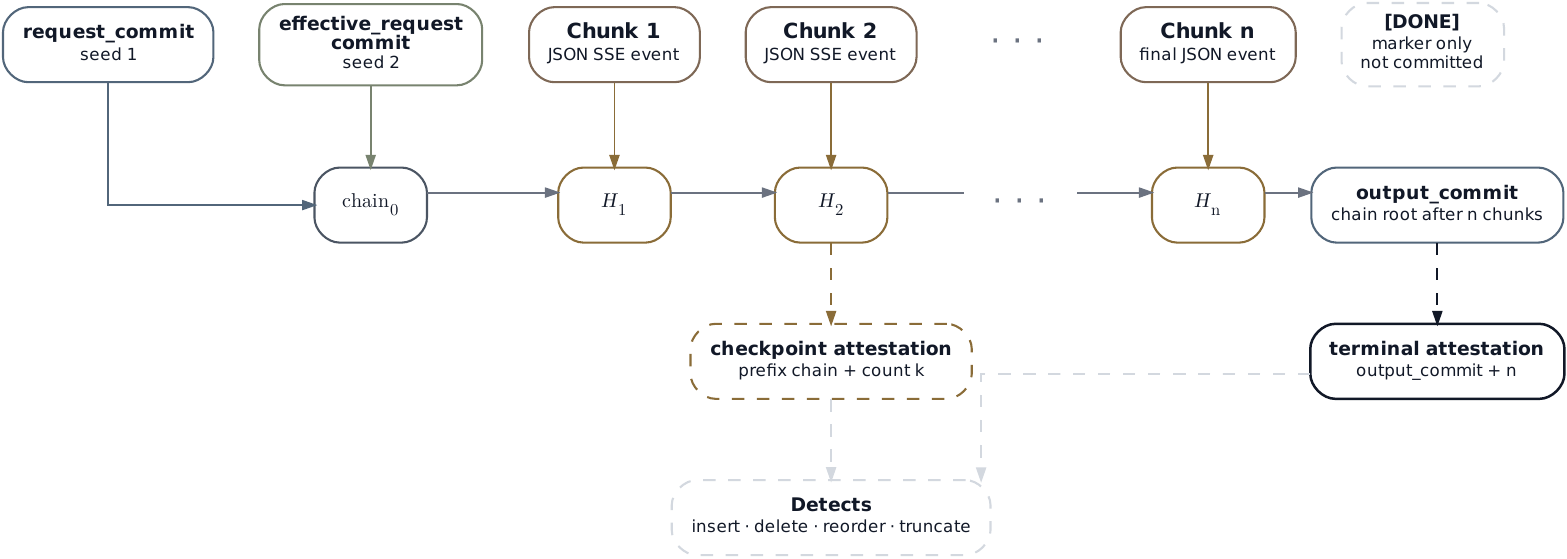}
\caption{Source-stream prefix mode in \AEX. Request anchors seed the source stream chain, each committed JSON event contributes to the complete source-stream commitment, checkpoints may certify verified prefixes, and the terminal attestation seals the complete stream commitment. Protocol markers such as \texttt{[DONE]} remain outside the chain.}
\label{fig:stream-prefix-mode}
\end{figure}

\begin{figure}[t]
\centering
\includegraphics[width=0.9\linewidth]{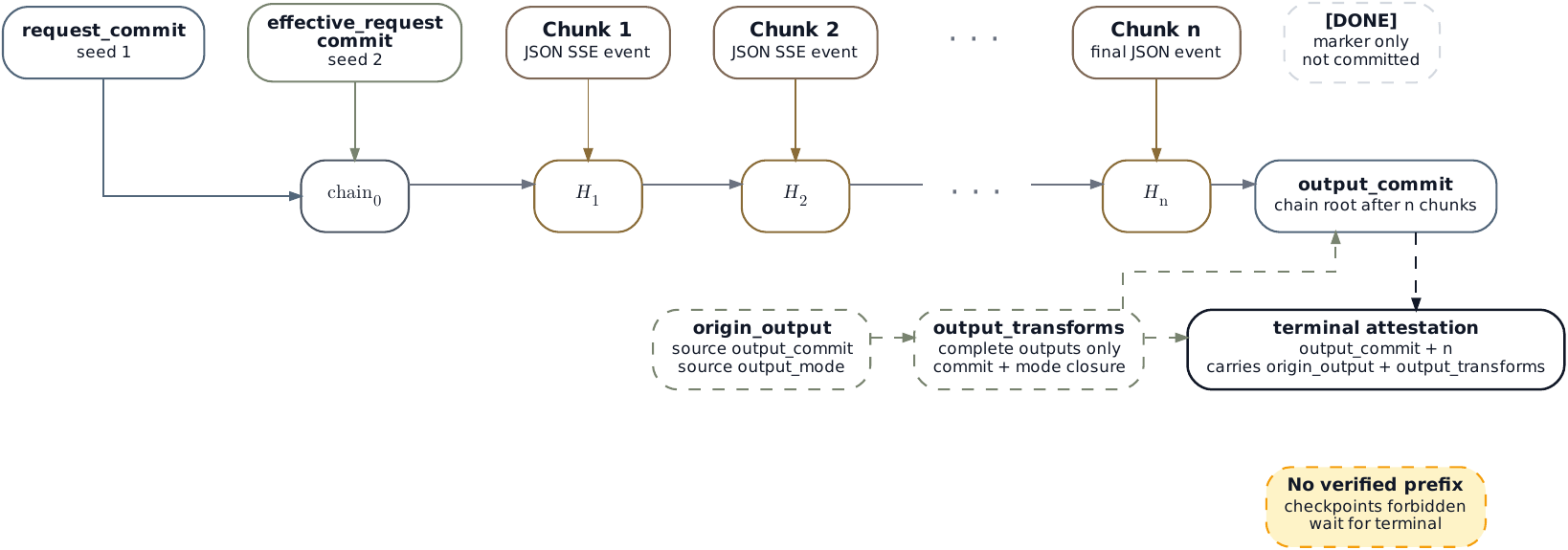}
\caption{Complete-output-lineage mode in \AEX. The delivered stream still has a request-anchored complete-output commitment, but checkpoints are forbidden and no verified prefix is exposed before the terminal attestation. Instead, the terminal attestation carries \texttt{origin\_output} and \texttt{output\_transforms}, which link the final delivered stream to a source complete output through a signed lineage that closes on both commitment values and \texttt{output\_mode}.}
\label{fig:stream-lineage-mode}
\end{figure}

Figures~\ref{fig:stream-prefix-mode} and~\ref{fig:stream-lineage-mode} contrast the two streaming proof modes. In both modes, request anchors seed the delivered stream's complete-output commitment. Source-stream prefix mode additionally allows checkpoints over an untransformed source stream, whereas complete-output-lineage mode forbids checkpoints and instead carries \texttt{origin\_output} plus \texttt{output\_transforms} only on the terminal attestation.

\subsection{Streaming proof modes}
\AEX\ supports exactly two streaming proof modes, and they are intentionally not mixed.

In \emph{source-stream prefix mode}, the delivered stream is the source issuer's own stream. A checkpoint attestation covers the prefix ending at chunk $k$ and includes the prefix chain value and chunk count $k$. The stream must still end with exactly one terminal attestation on the final JSON chunk. This mode lets a verifier distinguish ``verified prefix then truncation'' from ``no verifiable prefix at all.''

In \emph{complete-output-lineage mode}, the final delivered stream is treated as one complete output that may have descended from a different source complete output. The terminal attestation may therefore carry \texttt{origin\_output} and \texttt{output\_transforms} receipts instead of offering prefix proofs. In this mode, checkpoints are forbidden. A checkpoint attests only a source-stream prefix chain value; it is not itself a complete-output commitment and must never be reused as an \texttt{origin\_output} node or an \texttt{output\_transforms} endpoint.

\subsection{Request-transform receipts}
If a trusted intermediary rewrites the request, \AEX\ can include a sequence of request-transform receipts in the terminal attestation. Each receipt names an issuer, an input request commitment, an output request commitment, an algorithm, a key identifier, and a signature over the receipt body. The transform chain must begin at the original \texttt{request\_commit} and end at the attested \texttt{effective\_request\_commit}. Just as importantly, every step remains in the same commitment space. The same client-selected binding mode and the same nonce, if any, apply throughout. A transform chain can therefore explain trusted rewriting, but it cannot silently downgrade the original request-binding rule.

Operationally, these receipts are control-plane inputs to the final issuer rather than ordinary client payload. They should therefore travel on trusted intermediary-to-issuer channels such as proxy metadata or service-mesh context, not as unauthenticated client instructions that the issuer blindly echoes.

\begin{figure}[t]
\centering
\includegraphics[width=0.9\linewidth]{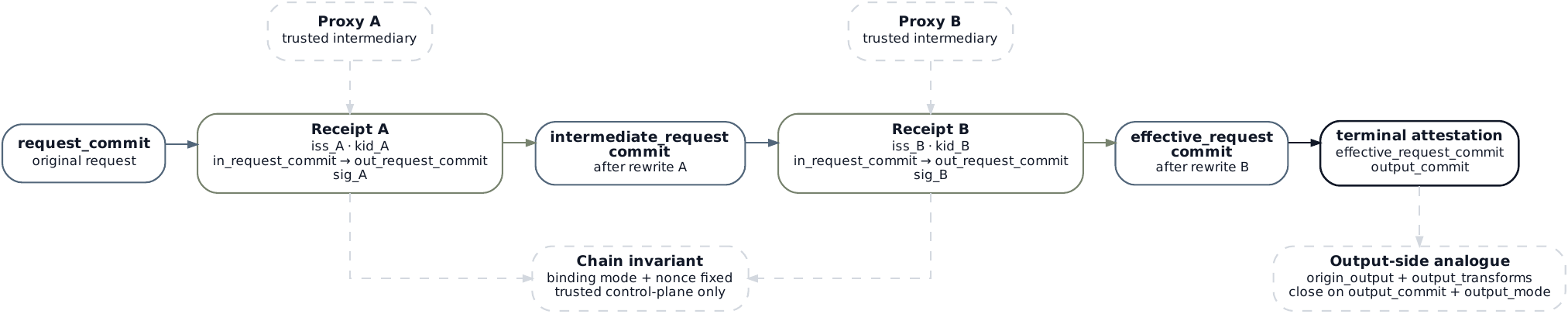}
\caption{Illustrative trusted request-transform chain. Each receipt is signed by a named intermediary, maps one request commitment into the next, and ultimately explains how the original \texttt{request\_commit} becomes the attested \texttt{effective\_request\_commit}. The binding mode and nonce remain fixed across the chain, the receipts are conveyed on trusted control-plane channels rather than as ordinary client payload, and the note at right highlights the analogous output-side closure pattern over \texttt{origin\_output}, \texttt{output\_transforms}, \texttt{output\_commit}, and \texttt{output\_mode}.}
\label{fig:transform-chain}
\end{figure}

Figure~\ref{fig:transform-chain} shows this more concretely: each trusted intermediary signs one receipt that maps the prior request commitment into the next commitment, while the binding mode and nonce stay fixed across the entire chain.

\subsection{Origin-output receipts and output-transform receipts}
If a trusted intermediary rewrites outputs, \AEX\ can include an \texttt{origin\_output} receipt and a sequence of \texttt{output\_transforms} receipts in the terminal attestation. The \texttt{origin\_output} receipt states that a source issuer signed an original complete-output commitment for the same request context. Each output-transform receipt then maps one complete-output commitment into the next, while carrying both an input \texttt{output\_mode} and an output \texttt{output\_mode}. The chain must begin at \texttt{origin\_output.output\_commit} and end at the terminal attestation's \texttt{output\_commit}, while closing on both commitment values and output modes.

These receipts are signed objects in their own right, using domain-separated payloads analogous to request-transform receipts. They are also control-plane metadata rather than client instructions. The terminal issuer must therefore verify them before re-signing them inside the top-level terminal attestation. A verifier need not be able to reconstruct every hidden intermediate output object to validate the lineage claim: the relevant claim is that trusted issuers signed the lineage edges and that the edges close on the final delivered complete output.

\subsection{Issuer discovery and trust}
\AEX\ assumes issuer-based key discovery using a JSON Web Key Set (JWKS) published at a well-known issuer-relative URL, concretely \texttt{\{iss\}/.well-known/aex-keys.json} \cite{rfc7517,rfc8615,rfc8414}. This choice is operational rather than cryptographic: many web-scale systems already use issuer + key-ID conventions. However, successful key lookup is not sufficient for trust. The verifier must still decide which issuers it trusts, for example through an allowlist, an origin-binding rule that matches the target API origin, or a comparable local policy. This requirement mirrors the broader remote-attestation literature, where attestation only has value relative to a verifier's trust policy \cite{menetrey2022attestation,akama2024rawebs}.

These same trust and discovery rules apply not only to the terminal attestation issuer but also to any request-transform issuer, source-output issuer, or output-transform issuer. To make the operational story less brittle, attestation objects may also carry issuance and expiry times, and verifiers should honor normal HTTP caching semantics when fetching JWKS documents \cite{rfc9111}. On a key-ID miss, refreshes should be rate-limited per issuer rather than retried unboundedly; during key rotation, old public keys should remain available long enough to validate attestations still within policy or expiry windows.

\section{Verification Model and Failure Semantics}
A useful attestation system must do more than answer ``valid'' or ``invalid.'' Streaming clients, auditors, and SDKs need structured failure semantics, and they need those semantics to respect the difference between source-stream prefix proofs and complete-output lineage proofs.

\subsection{Non-streaming verification}
A verifier processing a non-streaming response performs the following steps. Implementations may precompute local commitments for efficiency, but a response is accepted only if signature validation, issuer-trust checks, commitment comparisons, and any transform-chain verification all succeed.
\begin{enumerate}
    \item reconstruct the local request commitment from the original request, binding mode, and optional nonce;
    \item extract and remove the response's top-level \texttt{attestation};
    \item recompute the local complete-output commitment from the business payload and check that the terminal \texttt{output\_mode} is \texttt{non\_stream};
    \item verify the top-level attestation signature using the issuer's public key and local trust policy;
    \item if present, verify the request-transform chain and the effective-request commitment;
    \item if present, verify the \texttt{origin\_output} receipt, the \texttt{output\_transforms} chain, and closure of both commitment values and output modes;
    \item compare the locally reconstructed commitments to the attested commitments.
\end{enumerate}

The verifier need not be able to reconstruct any hidden source output or hidden intermediate output objects named only through \texttt{origin\_output} or \texttt{output\_transforms}. Those objects are validated through signature checks, request-context consistency checks, output-mode checks, and chain closure, not by requiring the client to recompute every hidden node.

\subsection{Streaming verification}
Streaming verification differs in one key way: chunk hashes may be accumulated before the verifier knows which proof mode the stream will end up using or whether the stream will end cleanly. The verifier therefore tracks chunk hashes, chunk counts, and attestation-bearing chunks as the stream arrives.

If the stream uses \emph{source-stream prefix mode}, a successful checkpoint yields a temporary \emph{verified prefix} state. The final terminal attestation must appear on the last JSON chunk, must carry \texttt{output\_mode = stream}, and must not carry \texttt{origin\_output} or \texttt{output\_transforms}. Missing or invalid terminal attestations then produce truncation or tampering states depending on whether a valid checkpoint was previously observed.

If the stream uses \emph{complete-output-lineage mode}, the verifier still accumulates the streaming chain over the delivered chunk sequence, but it must not interpret any received prefix as already verified output. Instead, it waits for the final terminal attestation, verifies the delivered stream's complete-output commitment, and then verifies the embedded \texttt{origin\_output} and \texttt{output\_transforms} lineage. In this mode, checkpoints are illegal.

\subsection{Recommended states}
We recommend the following core states:
\begin{itemize}
    \item \textbf{verified\_complete}: the request, output, signature, trust, and any transform-lineage checks all succeed;
    \item \textbf{verified\_prefix}: at least one checkpoint attestation verified, but no terminal attestation has been seen yet;
    \item \textbf{truncated\_after\_verified\_prefix}: a verified checkpoint exists, but the terminal attestation is missing or unusable;
    \item \textbf{truncated\_without\_terminal}: streamed chunks were seen, but the stream ended without a valid terminal attestation;
    \item \textbf{unattested\_or\_out\_of\_scope}: no applicable \AEX\ object is present, or the response lies outside the profile boundary under local policy;
    \item \textbf{request\_mismatch}: the original request commitment does not match the attested request commitment;
    \item \textbf{key\_unavailable}: the correct key could not be retrieved or resolved under local policy;
    \item \textbf{tampered}: a hard failure such as signature failure, output mismatch, illegal mixing of checkpoints with output transforms, issuer mismatch, transform-chain failure, or illegal attestation structure.
\end{itemize}

\textbf{verified\_prefix} and \textbf{truncated\_after\_verified\_prefix} apply only to source-stream prefix mode. In complete-output-lineage mode, a verifier should not expose any received prefix as already attested output, because the protocol intentionally does not define prefix-to-prefix output-transform proofs.

These states make explicit what many deployed clients lack: a machine-readable difference between a source stream that ended early after a verified prefix, a transformed stream that never delivered its final lineage proof, and a response that was never attested at all.

\begin{figure}[t]
\centering
\begin{tikzpicture}
\node[inner sep=0, anchor=south west] (img) at (0,0) {\includegraphics[width=0.9\linewidth]{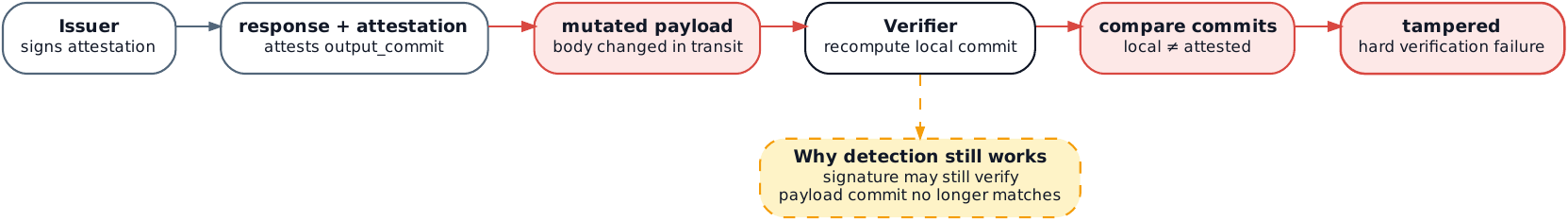}};
\begin{scope}[
    shift={(img.south west)},
    x={($(img.south east)-(img.south west)$)},
    y={($(img.north west)-(img.south west)$)}
]
\node[anchor=south east, inner sep=0, xshift=0.45mm, yshift=-1.0mm] at (0.489,0.662) {\includegraphics[width=4.4mm]{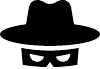}};
\end{scope}
\end{tikzpicture}
\caption{Illustrative tampering attempt and detection path. The issuer emits a signed response, an attacker mutates the payload in transit, and the verifier rejects the result as \texttt{tampered} because the locally recomputed payload commitment no longer matches the attested commitment, even if the detached signature object itself still parses or verifies.}
\label{fig:tamper-detection}
\end{figure}

Figure~\ref{fig:tamper-detection} illustrates a representative \texttt{tampered} outcome: an attacker alters a signed response after issuance, the verifier recomputes the local payload commitment, and the mismatch is surfaced explicitly rather than being silently accepted.

\section{Security and Privacy Analysis}
\subsection{What \AEX\ proves}
\AEX\ proves that a trusted issuer signed a specific relation between a request commitment and an output commitment. If \texttt{effective\_request\_commit} and \texttt{request\_transforms} are present, it additionally proves that the issuer signed a relation from the original request commitment, through an explicit chain of trusted request transforms, to the effective request commitment, and then to the output.

If \texttt{origin\_output} and \texttt{output\_transforms} are present, \AEX\ additionally proves that a source issuer signed an original complete-output commitment for the same request context and that a continuous, signed chain of trusted output transforms closes from that source commitment to the final complete-output commitment delivered to the client. In source-stream prefix mode, checkpoints additionally prove that a source issuer signed a particular untransformed stream prefix.

This is a narrower but more realistic claim than proving exact model identity. Fingerprinting and MET are still useful for detecting that an API appears to serve a different model or output distribution \cite{pasquini2025llmmap,gao2025met}; \AEX\ instead provides a signed transaction binding and, when needed, a signed complete-output lineage.

\subsection{What \AEX\ does not prove}
\AEX\ should not be read as evidence of exact model identity or hidden-behavior absence. Those concerns require different evidence channels, such as model-auditing methods, TEEs, or provider disclosure \cite{zhang2026shadow,sun2025svip,zhang2025attestllm}. Nor does \AEX\ prove that every hidden intermediate output in an accepted output-transform chain is visible to the client, or that an accepted transform is semantically correct, reasonable, or minimal.

\subsection{Replay and nonce use}
Without a nonce, request commitments are deterministic. For low-entropy requests, an observer who sees an attestation may be able to guess the underlying request by dictionary enumeration. Client-supplied nonces reduce this linkability and can support replay detection when combined with freshness windows, issuance times, or verifier-side replay policy. For backward compatibility, \AEX\ does not make nonces globally mandatory, but deployed profiles should strongly encourage high-entropy client nonces whenever request privacy or freshness matters, especially when complete-output lineage metadata is returned. The design is similar in spirit to replay mitigations discussed in HTTP Message Signatures, which note the importance of choosing signed components and nonces carefully to avoid replayable signatures \cite{rfc9421}.

\subsection{Trusted rewriting versus silent mutation}
Silent request mutation is a serious problem for LLM APIs. Even if an endpoint really forwards a request to the expected model family, undeclared prompt preprocessing, context truncation, or inference-parameter changes can alter behavior substantially \cite{zhang2026shadow}. By separating original and effective request commitments and making accepted rewrites explicit, \AEX\ prevents the verifier from having to choose between two bad extremes: treating all mutation as harmless noise, or treating all mutation as fatal mismatch even when it is documented and trusted.

\subsection{Trusted output rewriting and complete-output lineage}
Output-side mutation presents an analogous problem. A gateway may legitimately redact sensitive spans, normalize tool-call envelopes, aggregate a stream into one final object, or re-package one complete object as a stream. Treating every such change as simple tampering would make the protocol unusable for many real deployments; treating every such change as harmless would erase provenance. By separating a source complete output from the final complete output, and by requiring signed \texttt{origin\_output} and \texttt{output\_transforms} receipts, \AEX\ lets verifiers distinguish trusted, explicit output rewriting from silent mutation.

\subsection{Cross-mode output transforms}
\AEX\ permits output lineage to cross between non-streaming and streaming forms. This is operationally realistic, but it requires explicit \texttt{output\_mode} values on every lineage node and mode closure along the chain. Otherwise a verifier could accidentally compare commitments that were computed over different objects under different algorithms and accept an invalid lineage. The explicit mode labels are therefore a security boundary, not just descriptive metadata.

\subsection{Checkpoints, truncation, and midstream tampering}
The streaming chain protects against content mutation, insertion, deletion, reordering, and truncation. Classical stream-authentication work observes that authenticating a stream efficiently requires linking data units together so that integrity failures are detectable without signing each unit independently \cite{golle2001streamauth}. \AEX\ applies this lesson to JSON chunk streams and augments it with request anchors and explicit terminal semantics.

\subsection{Why checkpoints and output transforms are mutually exclusive}
The mutual exclusion rule between checkpoints and \texttt{output\_transforms} is deliberate. A checkpoint proves something about a source issuer's \emph{original stream prefix}. Once output transforms are introduced, the final delivered output is no longer the same object as that source prefix stream. \AEX\ v1 therefore avoids a misleading stronger claim. It keeps checkpoints as source-prefix proofs, keeps output transforms as complete-output lineage proofs, and refuses to define prefix-to-prefix lineage across rewritten outputs in this version.

\subsection{Issuer trust and key-discovery abuse}
A verifier must not trust an issuer merely because that issuer serves a valid JWKS \cite{rfc7517}. Issuer trust must be established by local policy, target-origin matching, pinning, or a comparable external trust rule. This applies to terminal issuers, request-transform issuers, source-output issuers, and output-transform issuers alike. In addition, unknown key identifiers can be abused to provoke repeated JWKS refreshes. A robust implementation should therefore rate-limit bypass-cache JWKS refresh attempts per issuer.

\subsection{Privacy implications of commitments and lineage metadata}
Commitment values are deterministic: the same committed object yields the same commitment. That means low-entropy requests or outputs may be guessable by dictionary attack. Output-lineage metadata can also reveal structural facts even when the hidden intermediate objects are not disclosed. For example, \texttt{origin\_output} and \texttt{output\_transforms} may reveal that the source and final outputs differ, which trusted intermediaries participated, which transform policy labels were used, and whether the lineage crossed between non-streaming and streaming forms. Archived checkpoints may reveal that a source issuer had already signed a particular stream prefix at a particular time. Deployments that log or publish attestations should therefore evaluate carefully which commitment values, receipts, and policy annotations they retain or expose.

\subsection{Interaction with TEEs}
TEE remote attestation proves something different from \AEX. It helps establish trust in the serving environment and measured code, while \AEX\ binds application-layer request and response objects, streaming chunks, and when needed complete-output lineage metadata. The two mechanisms are complementary. A TEE may act as the issuer; equally, an \AEX\ issuer may exist without a TEE. The distinction is important because remote attestation alone does not define how client-visible API objects, streaming chunks, or trusted middleware transforms are represented at the application layer \cite{akama2024rawebs,menetrey2022attestation}.

\section{Related Work and Positioning}
\subsection{HTTP message signing and JSON canonicalization}
The closest standard-adjacent ancestor to \AEX\ is HTTP Message Signatures \cite{rfc9421}. RFC 9421 defines how to sign semantically meaningful HTTP components under transformation, supports signing request components in a response, and explicitly acknowledges intermediary-induced transformation \cite{rfc9421}. \AEX\ borrows that mindset but specializes it to JSON LLM APIs. The specialization matters because \AEX\ adds JSON-object commitments, request-binding modes, explicit absent-field binding, chunk-stream semantics, request-transform receipts, and source-output / output-transform lineage receipts.

At the representation layer, \AEX\ depends on JCS to canonicalize JSON data for stable hashing and signing \cite{rfc8785}. This is a pragmatic design choice: the protocol seeks interoperability over existing JSON objects rather than inventing a new serialization.

\subsection{Supply-chain and media provenance}
in-toto and DSSE provide a useful conceptual baseline. in-toto secures software supply chains by chaining signed metadata across steps, while DSSE provides a simple envelope for signing arbitrary payloads \cite{torresarias2019intoto,dsse}. \AEX\ shares the idea that provenance often needs to describe \emph{relations} between states rather than sign a single blob in isolation. However, software-supply-chain attestations are offline and artifact-centric, whereas \AEX\ targets online LLM API transactions. C2PA likewise offers strong provenance machinery for digital media and content credentials, but its focus is media-asset lineage rather than interactive request/stream semantics \cite{c2pa-spec}.

\subsection{Auditing and fingerprinting LLM APIs}
LLMmap and MET study a different problem: given an endpoint, what evidence can a user collect to infer which model or distribution is being served? LLMmap uses active fingerprinting to identify likely models across unknown prompts and application layers \cite{pasquini2025llmmap}. MET formalizes black-box endpoint auditing as a two-sample testing problem and shows that many inference endpoints deviate from reference distributions \cite{gao2025met}. Zhang et al. apply related tools in their shadow-API audit and find both identity mismatches and behavior drift \cite{zhang2026shadow}. \AEX\ complements this line of work. Rather than infer model identity from outputs, it binds a request-output transaction to a trusted issuer.

\subsection{Verifiable inference and trusted execution}
SVIP and AttestLLM attack model-verification problems more directly. SVIP proposes a secret-based scheme that uses hidden representations as model identifiers for open-source LLM inference \cite{sun2025svip}. AttestLLM focuses on billion-scale on-device LLMs and combines hardware/software techniques with a TEE-oriented attestation workflow \cite{zhang2025attestllm}. RA-WEBs, meanwhile, designs a web-compatible remote-attestation protocol for services using TEEs \cite{akama2024rawebs}. These approaches are valuable but answer a stronger and often more intrusive question: what code or model actually ran? \AEX\ is intentionally weaker and more deployable. It can operate with black-box APIs and existing request/response formats.

\begin{table*}[t]
\centering
\caption{Positioning \AEX\ against neighboring mechanisms. The table summarizes the design space rather than claiming mutual exclusivity.}
\label{tab:positioning}
\begin{tabularx}{\textwidth}{>{\raggedright\arraybackslash}p{2.7cm}>{\raggedright\arraybackslash}p{2.2cm}>{\raggedright\arraybackslash}p{2.4cm}>{\raggedright\arraybackslash}p{2.3cm}>{\raggedright\arraybackslash}X}
\toprule
Mechanism & Main object & Online API transaction binding & Streaming chunk semantics & Primary question answered \\
\midrule
HTTP Message Signatures \cite{rfc9421} & HTTP components & Partial, generic & No LLM-specific chunk model & Which HTTP components were signed, despite safe transformation? \\
C2PA \cite{c2pa-spec} & Media assets & No & No & What is the provenance history of this media asset? \\
in-toto / DSSE \cite{torresarias2019intoto,dsse} & Supply-chain metadata & Not transaction-oriented & No & What signed steps and artifacts formed a software supply chain? \\
LLMmap / MET \cite{pasquini2025llmmap,gao2025met} & Output behavior & Inferential only & Indirect & Which model or distribution does this endpoint appear to serve? \\
SVIP \cite{sun2025svip} & Model internals + outputs & Yes, but intrusive & Not targeted at API chunk streams & Did the provider likely use the requested open-source model? \\
TEE attestation / RA-WEBs / AttestLLM \cite{menetrey2022attestation,akama2024rawebs,zhang2025attestllm} & Execution environment & Indirect unless coupled with app semantics & No LLM-API semantics by default & What measured code/environment is running? \\
\textbf{\AEX\ (this paper)} & JSON requests, complete outputs, and lineage receipts & \textbf{Yes} & \textbf{Yes} & Which issuer attested to this request-output relation or complete-output lineage at the API boundary? \\
\bottomrule
\end{tabularx}
\end{table*}

\section{Deployment and Standardization Path}
\subsection{Why a protocol extension?}
The shadow-API literature points toward a need for lightweight, official verification endpoints \cite{zhang2026shadow}. A practical verification endpoint should require minimal application changes, work with existing API formats, and compose with SDKs and gateways. \AEX\ is designed to fit that niche. Because it adds only a top-level \texttt{attestation} object and leaves the rest of the payload untouched, existing clients can adopt it incrementally.

\subsection{OpenAI-compatible profile}
The most obvious first profile is an OpenAI-compatible chat-completions interface. OpenAI's chat-completions endpoint accepts a list of messages and can return either a full chat-completion object or a streamed sequence of chunk objects \cite{openai-chat,openai-chat-stream}. \AEX\ can attach to that boundary directly, making it possible to support existing request shapes, multimodal message content, and tool-calling payloads without inventing a new application protocol.

For streaming, the profile treats each complete JSON SSE event as one committed chunk and leaves protocol markers such as \texttt{[DONE]} outside the chain \cite{whatwg-sse,openai-chat-stream}. A practical deployment can request streaming attestations with a profile-level control such as \texttt{stream\_options.include\_attestation}. Business fields such as \texttt{delta.content}, \texttt{delta.refusal}, and \texttt{delta.tool\_calls} remain unchanged and are covered simply because they are part of the committed chunk objects.

The profile naturally supports both \AEX\ streaming proof modes. In source-stream prefix mode, checkpoints may appear on intermediate JSON chunks and the final terminal attestation must not carry \texttt{origin\_output} or \texttt{output\_transforms}. In complete-output-lineage mode, checkpoints do not appear; instead, the final terminal attestation may carry \texttt{origin\_output} and \texttt{output\_transforms}, and the source \texttt{output\_mode} may differ from the final delivered \texttt{output\_mode}. The same profile also permits non-streaming responses to carry output-lineage metadata, for example when a gateway buffers a source stream into one final JSON object before returning it.

If the profile represents errors as top-level JSON objects, those errors can be attested in the same way as ordinary non-streaming responses.

\subsection{Implementation patterns}
Several deployment patterns are plausible.
\begin{itemize}
    \item \textbf{Origin signing:} the provider signs responses at the serving edge.
    \item \textbf{Gateway signing:} a trusted gateway computes commitments and signs them.
    \item \textbf{TEE-assisted signing:} a confidential-computing environment or enclave-backed service emits the attestation, possibly after separate remote-attestation checks \cite{akama2024rawebs,menetrey2022attestation}.
    \item \textbf{Auditor verification:} an SDK or proxy verifies signatures and surfaces structured states to applications.
\end{itemize}

\subsection{What remains to standardize}
This paper presents the protocol core and an initial profile. A standardization path would still need to define key-discovery metadata, profile registries, and implementation guidance for gateways and SDKs. It would also benefit from cross-language test vectors for canonicalization, request-binding projections, and streaming event boundaries, together with a public verifier library and independent implementations. Those are engineering and standardization tasks rather than open cryptographic questions.

\section{Reference Prototype, Experiments, and Testing}
\label{sec:prototype-eval}
\subsection{Prototype and testbed}
To validate that the protocol can be implemented without changing application-visible API semantics, we built a reference TypeScript prototype in the repository accompanying this preprint. The prototype has three parts: a core library that implements canonicalization, signing, verification, and JWKS caching; a standalone lab service that exposes an OpenAI-compatible chat-completions endpoint; and an optional browser workbench that consumes the service's public API but is not part of the trust boundary.

The lab service is a single-host testbed with one HTTP gateway, two internal HTTPS proxies, and one internal HTTPS provider simulator. Proxy A, Proxy B, and the provider each act as issuers and publish issuer-relative JWKS documents. The gateway acts as the verifier and stores a per-run trace that includes the request text, any effective request after trusted rewriting, raw SSE events, verifier states, and structured diagnostics. The artifact exercises the request-binding modes, effective-request commitments, request-transform chains, streaming checkpoints, complete-output-lineage verification via \texttt{origin\_output} and \texttt{output\_transforms}, cross-mode output rewriting, and negative-path failure states discussed earlier in this paper. It is intentionally a protocol testbed rather than a model-quality benchmark: the upstream component is a deterministic simulator, not a production LLM service.

\begin{figure}[t]
\centering
\includegraphics[width=0.9\linewidth]{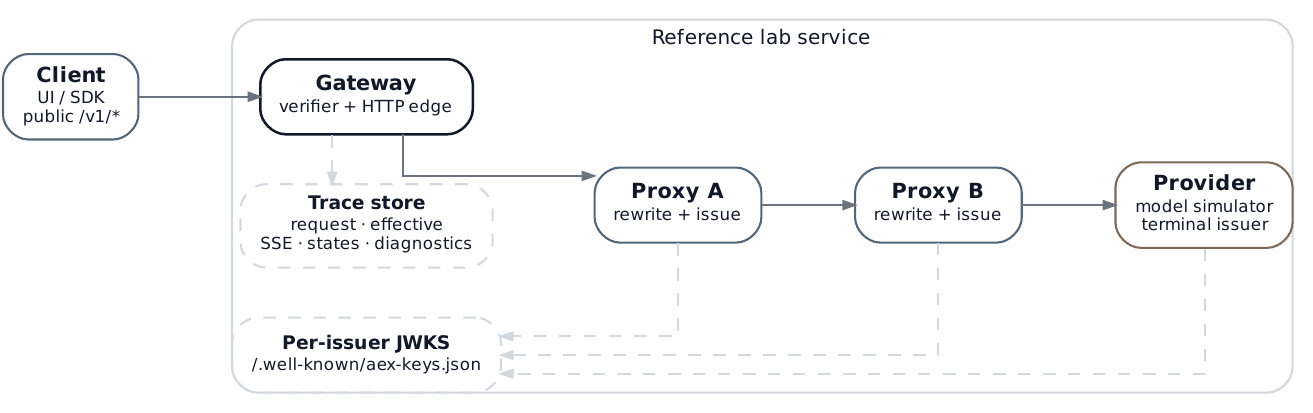}
\caption{Reference prototype topology used in the accompanying artifact. The public client surface terminates at the gateway verifier; the internal proxies model trusted request rewriting while the provider emits source-stream and terminal evidence; and the gateway stores a per-run trace with request, effective-request, SSE, checkpoint, and verifier-state diagnostics.}
\label{fig:lab-topology}
\end{figure}

Figure~\ref{fig:lab-topology} shows the concrete topology used in the artifact and clarifies which components are inside the signed request/response path versus which are merely consumers of the public API.

\subsection{Automated validation}
We validated the prototype with unit, integration, and browser end-to-end tests. Table~\ref{tab:validation} keeps only the top-level validation counts, and all listed tests passed in the validation rerun reported here. The unit tests cover request-binding projections, non-stream attestation issuance and verification, stream checkpoint and terminal verification, non-stream output-lineage receipts, stream output-lineage verification, and JWKS cache behavior. The integration tests exercise the full gateway/proxy/provider stack, including checkpoint-mode streaming, non-stream lineage, stream lineage, and negative-path states such as \texttt{request\_mismatch}, \texttt{tampered}, and truncated streams. The browser tests verify that the public \texttt{/v1/*} service surface drives the workbench correctly. We additionally ran a repository-wide TypeScript build check and an automated traceability-matrix consistency check; both passed, with the traceability script validating 13 rows, 58 references, and 11 required section anchors.

\begin{table}[t]
\centering
\caption{Automated validation in the reference prototype.}
\label{tab:validation}
\begin{tabularx}{\linewidth}{>{\raggedright\arraybackslash}p{2.4cm}>{\centering\arraybackslash}p{1.3cm}X}
\toprule
Layer & Count & Main coverage \\
\midrule
Unit tests & 29 & Core request binding, attestation issuance/verification, output-lineage receipts, stream verification, and JWKS behavior \\
Integration tests & 25 & End-to-end gateway/proxy/provider flows, request/output transform chains, lineage scenarios, and negative-path verification states \\
Browser end-to-end tests & 6 & Public \texttt{/v1/*} API wiring, stream progress, diagnostics, and proxy configuration workflows \\
\bottomrule
\end{tabularx}
\end{table}

\subsection{Local microbenchmarks}
We also ran a small local benchmark to sanity-check overhead in the reference implementation. The benchmark script starts an isolated copy of the lab stack on one Ubuntu~24.04 x86-64 host, performs 3 warmup runs, and then records 20 measured runs for each of five success scenarios covering non-stream success, checkpoint-mode streaming, request-transform-chain success, non-stream lineage, and stream lineage. For each run, it records client-side latency as observed by the benchmark driver and server-side timing, size, and event counters from the gateway's stored run trace.

These measurements should be interpreted narrowly. First, they are single-host measurements on a deterministic simulator rather than on a remote production deployment. Second, the streaming scenarios intentionally insert two fixed 120~ms delays between emitted chunks in order to make checkpoint and terminal behavior visible; the absolute end-to-end stream latency is therefore dominated by those inserted delays rather than by cryptographic work. Third, in this implementation the JWKS cache is scoped to a verification run, so the request-transform-chain scenario performs three JWKS fetches (Proxy A, Proxy B, and Provider) by construction, whereas the lineage scenarios perform two. Fourth, the table below still reports only five representative success cases rather than every implemented lineage configuration, so it should be read as a sanity check on local overhead rather than as a full performance characterization of all protocol paths.

Table~\ref{tab:microbench} reports only the median latency figures that are most useful for interpreting the prototype. Across all 100 measured runs, the five benchmarked scenarios returned HTTP 200 and completed with \texttt{verified\_complete}. In the simple non-streaming case, median client-observed completion was 2.287~ms and median gateway verification time was 0.495~ms. In the source-stream checkpoint case, median time to the first streamed chunk was 1.736~ms, while full completion was 243.371~ms because of the simulator's fixed delays; the gateway's own median verification time remained 0.843~ms. In the trusted request-transform-chain case, median gateway verification time rose to 0.952~ms, with two transform receipts and three JWKS fetches per run. In the non-stream lineage case, median client-observed completion was 2.225~ms and median gateway verification time was 0.710~ms. In the stream-lineage case, median time to the first streamed chunk was 1.666~ms, full completion was 243.028~ms, and median gateway verification time was 1.491~ms; both lineage scenarios carried one transform receipt and two JWKS fetches per run in the reference implementation.

\begin{table}[t]
\centering
\small
\caption{Local microbenchmarks on the reference lab stack. The table reports only the median latency figures used in the discussion text.}
\label{tab:microbench}
\begin{tabularx}{\linewidth}{>{\raggedright\arraybackslash}p{2.8cm}>{\centering\arraybackslash}p{1.4cm}>{\centering\arraybackslash}p{1.9cm}>{\centering\arraybackslash}p{1.9cm}>{\centering\arraybackslash}X}
\toprule
Scenario & Runs & Median client completion (ms) & Median first stream chunk (ms) & Median gateway verification (ms) \\
\midrule
Non-stream success & 20/20 verified & 2.287 & --- & 0.495 \\
Stream success (checkpoint) & 20/20 verified & 243.371 & 1.736 & 0.843 \\
Transform-chain success & 20/20 verified & 2.609 & --- & 0.952 \\
Non-stream lineage success & 20/20 verified & 2.225 & --- & 0.710 \\
Stream lineage success & 20/20 verified & 243.028 & 1.666 & 1.491 \\
\bottomrule
\end{tabularx}
\end{table}

\section{Limitations and Open Questions}
This paper remains primarily a design and analysis preprint, although Section~\ref{sec:prototype-eval} reports a reference prototype and local measurements. Several limitations are therefore explicit.

\begin{itemize}
    \item \textbf{Only local prototype measurements in this version.} Section~\ref{sec:prototype-eval} reports single-host measurements from a simulator-based reference implementation. These results are useful as a sanity check on implementability and local overhead, but they are not cross-provider interoperability results, WAN measurements, or production capacity benchmarks.
    \item \textbf{Benchmark coverage remains selective.} The repository prototype exercises detached terminal attestations, request-binding modes, request-transform chains, streaming checkpoints, and the broader \texttt{origin\_output} / \texttt{output\_transforms} lineage model end-to-end. However, the microbenchmark table in Section~\ref{sec:prototype-eval} still reports only five representative success scenarios rather than every implemented lineage configuration.
    \item \textbf{No public cross-language interop suite.} This work does not ship a public verifier together with cross-language test vectors and independent implementations. Those will matter for production deployment and standardization.
    \item \textbf{No direct model-identity or hidden-behavior proof.} \AEX\ can coexist with fingerprinting, MET, or TEE attestation, but it does not replace them, nor can it prove the absence of hidden prompts, private retrieval, or internal policy engines \cite{pasquini2025llmmap,gao2025met,sun2025svip,zhang2025attestllm}.
    \item \textbf{Issuer trust remains foundational.} If the issuer is malicious or the verifier trusts the wrong issuer, \AEX\ does not rescue the system.
    \item \textbf{No prefix-to-prefix output-transform proofs in v1.} The protocol revision studied here intentionally refuses to translate a source-stream checkpoint into a verified prefix of a transformed output. Once output transforms are present, \AEX\ falls back to complete-output lineage only. Richer prefix-to-prefix lineage is a possible future extension, but it would materially increase protocol and verifier complexity.
    \item \textbf{Future profile work is needed.} This paper focuses on JSON objects and JSON-object streams. Binary responses, audio streams, and non-JSON protocols need separate treatment.
\end{itemize}

A particularly important open question is how to combine \AEX\ with stronger evidence sources. One attractive path is layering: a TEE-backed endpoint can first prove the serving environment using remote attestation and then use \AEX\ to bind that environment to the exact request and response objects seen by the application. Another path is policy composition: an auditor could combine an \AEX\ verification result with LLMmap or MET to distinguish ``issuer signed this transaction'' from ``issuer signed a transaction that still appears behaviorally inconsistent with the claimed model.''

\section{Conclusion}
Hosted LLM APIs expose an important trust boundary that remains under-specified by existing infrastructures. Recent evidence from shadow-API auditing shows that the boundary is not only theoretically interesting but practically fragile \cite{zhang2026shadow}. Existing techniques such as fingerprinting, statistical testing, verifiable inference, and remote attestation each illuminate part of the problem, but they do not directly provide a low-friction, transaction-bound attestation layer for existing APIs.

\AEX\ is a proposal for that missing layer. By attaching a signed top-level attestation to an otherwise ordinary JSON response, and by defining request commitments, effective request commitments, request-transform receipts, source-output receipts, output-transform receipts, and output-mode-aware streaming commitments, \AEX\ offers a practical path toward verifiable request-output binding and complete-output lineage at the LLM API boundary. It is intentionally scoped to that transaction boundary: a trusted issuer signed a specific request-output relation, and when applicable a trusted chain of issuers signed how the delivered complete output descended from a source output. That scope keeps \AEX\ deployable, composable, and well matched to the API-centric way LLM systems are actually consumed today.

\appendix
\section{Illustrative JSON Examples}
This appendix gives simplified examples of \AEX\ syntax. The digests and signatures shown are placeholders.

\subsection{Request with explicit binding mode}
\begin{lstlisting}
{
  "model": "gpt-5",
  "messages": [
    {"role": "user", "content": "Summarize this document."}
  ],
  "attestation": {
    "required": true,
    "nonce": "Q2xpZW50Tm9uY2UxMjM",
    "request_binding": {
      "mode": "top_level_include",
      "fields": ["messages", "model", "tools"]
    }
  }
}
\end{lstlisting}

\subsection{Non-streaming response}
\begin{lstlisting}
{
  "id": "chatcmpl-123",
  "choices": [
    {
      "index": 0,
      "message": {"role": "assistant", "content": "Here is the summary."},
      "finish_reason": "stop"
    }
  ],
  "attestation": {
    "version": "1",
    "kind": "terminal",
    "profile": "openai.chat_completions",
    "iss": "https://gateway.example",
    "request_commit": "sha256:aaaaaaaaaaaaaaaaaaaaaaaaaaaaaaaaaaaaaaaaaaaaaaaaaaaaaaaaaaaaaaaa",
    "effective_request_commit": "sha256:eeeeeeeeeeeeeeeeeeeeeeeeeeeeeeeeeeeeeeeeeeeeeeeeeeeeeeeeeeeeeeee",
    "request_transforms": [
      {
        "iss": "https://proxy-a.example",
        "in_request_commit": "sha256:aaaaaaaaaaaaaaaaaaaaaaaaaaaaaaaaaaaaaaaaaaaaaaaaaaaaaaaaaaaaaaaa",
        "out_request_commit": "sha256:eeeeeeeeeeeeeeeeeeeeeeeeeeeeeeeeeeeeeeeeeeeeeeeeeeeeeeeeeeeeeeee",
        "policy": "normalize-openai-chat/v1",
        "alg": "Ed25519",
        "kid": "proxy-a-key-2",
        "sig": "BASE64URL_SIGNATURE"
      }
    ],
    "output_mode": "non_stream",
    "origin_output": {
      "profile": "openai.chat_completions",
      "iss": "https://api.provider.com",
      "request_commit": "sha256:aaaaaaaaaaaaaaaaaaaaaaaaaaaaaaaaaaaaaaaaaaaaaaaaaaaaaaaaaaaaaaaa",
      "effective_request_commit": "sha256:eeeeeeeeeeeeeeeeeeeeeeeeeeeeeeeeeeeeeeeeeeeeeeeeeeeeeeeeeeeeeeee",
      "output_mode": "stream",
      "output_commit": "sha256:cccccccccccccccccccccccccccccccccccccccccccccccccccccccccccccccc",
      "alg": "Ed25519",
      "kid": "provider-key-1",
      "sig": "BASE64URL_SIGNATURE"
    },
    "output_transforms": [
      {
        "iss": "https://gateway.example",
        "request_commit": "sha256:aaaaaaaaaaaaaaaaaaaaaaaaaaaaaaaaaaaaaaaaaaaaaaaaaaaaaaaaaaaaaaaa",
        "effective_request_commit": "sha256:eeeeeeeeeeeeeeeeeeeeeeeeeeeeeeeeeeeeeeeeeeeeeeeeeeeeeeeeeeeeeeee",
        "in_output_mode": "stream",
        "in_output_commit": "sha256:cccccccccccccccccccccccccccccccccccccccccccccccccccccccccccccccc",
        "out_output_mode": "non_stream",
        "out_output_commit": "sha256:bbbbbbbbbbbbbbbbbbbbbbbbbbbbbbbbbbbbbbbbbbbbbbbbbbbbbbbbbbbbbbbb",
        "policy": "buffer-and-collapse/v1",
        "alg": "Ed25519",
        "kid": "gateway-key-9",
        "sig": "BASE64URL_SIGNATURE"
      }
    ],
    "output_commit": "sha256:bbbbbbbbbbbbbbbbbbbbbbbbbbbbbbbbbbbbbbbbbbbbbbbbbbbbbbbbbbbbbbbb",
    "nonce": "Q2xpZW50Tm9uY2UxMjM",
    "alg": "Ed25519",
    "kid": "gateway-key-9",
    "sig": "BASE64URL_SIGNATURE"
  }
}
\end{lstlisting}

\subsection{Terminal streaming chunk in source-stream prefix mode}
\begin{lstlisting}
{
  "id": "chatcmpl-123",
  "choices": [],
  "usage": {
    "prompt_tokens": 128,
    "completion_tokens": 42,
    "total_tokens": 170
  },
  "attestation": {
    "version": "1",
    "kind": "terminal",
    "profile": "openai.chat_completions",
    "iss": "https://api.example.com",
    "request_commit": "sha256:aaaaaaaaaaaaaaaaaaaaaaaaaaaaaaaaaaaaaaaaaaaaaaaaaaaaaaaaaaaaaaaa",
    "effective_request_commit": "sha256:eeeeeeeeeeeeeeeeeeeeeeeeeeeeeeeeeeeeeeeeeeeeeeeeeeeeeeeeeeeeeeee",
    "request_transforms": [
      {
        "iss": "https://proxy-a.example",
        "in_request_commit": "sha256:aaaaaaaaaaaaaaaaaaaaaaaaaaaaaaaaaaaaaaaaaaaaaaaaaaaaaaaaaaaaaaaa",
        "out_request_commit": "sha256:eeeeeeeeeeeeeeeeeeeeeeeeeeeeeeeeeeeeeeeeeeeeeeeeeeeeeeeeeeeeeeee",
        "policy": "normalize-openai-chat/v1",
        "alg": "Ed25519",
        "kid": "proxy-a-key-2",
        "sig": "BASE64URL_SIGNATURE"
      }
    ],
    "output_mode": "stream",
    "output_commit": "sha256:dddddddddddddddddddddddddddddddddddddddddddddddddddddddddddddddd",
    "chunk_count": "42",
    "alg": "Ed25519",
    "kid": "edge-key-2026-01",
    "sig": "BASE64URL_SIGNATURE"
  }
}
\end{lstlisting}

\subsection{Terminal streaming chunk in complete-output-lineage mode}
\begin{lstlisting}
{
  "id": "chatcmpl-123",
  "choices": [],
  "usage": {
    "prompt_tokens": 128,
    "completion_tokens": 42,
    "total_tokens": 170
  },
  "attestation": {
    "version": "1",
    "kind": "terminal",
    "profile": "openai.chat_completions",
    "iss": "https://gateway.example",
    "request_commit": "sha256:aaaaaaaaaaaaaaaaaaaaaaaaaaaaaaaaaaaaaaaaaaaaaaaaaaaaaaaaaaaaaaaa",
    "effective_request_commit": "sha256:eeeeeeeeeeeeeeeeeeeeeeeeeeeeeeeeeeeeeeeeeeeeeeeeeeeeeeeeeeeeeeee",
    "request_transforms": [
      {
        "iss": "https://proxy-a.example",
        "in_request_commit": "sha256:aaaaaaaaaaaaaaaaaaaaaaaaaaaaaaaaaaaaaaaaaaaaaaaaaaaaaaaaaaaaaaaa",
        "out_request_commit": "sha256:eeeeeeeeeeeeeeeeeeeeeeeeeeeeeeeeeeeeeeeeeeeeeeeeeeeeeeeeeeeeeeee",
        "policy": "normalize-openai-chat/v1",
        "alg": "Ed25519",
        "kid": "proxy-a-key-2",
        "sig": "BASE64URL_SIGNATURE"
      }
    ],
    "output_mode": "stream",
    "origin_output": {
      "profile": "openai.chat_completions",
      "iss": "https://api.provider.com",
      "request_commit": "sha256:aaaaaaaaaaaaaaaaaaaaaaaaaaaaaaaaaaaaaaaaaaaaaaaaaaaaaaaaaaaaaaaa",
      "effective_request_commit": "sha256:eeeeeeeeeeeeeeeeeeeeeeeeeeeeeeeeeeeeeeeeeeeeeeeeeeeeeeeeeeeeeeee",
      "output_mode": "non_stream",
      "output_commit": "sha256:9999999999999999999999999999999999999999999999999999999999999999",
      "alg": "Ed25519",
      "kid": "provider-key-1",
      "sig": "BASE64URL_SIGNATURE"
    },
    "output_transforms": [
      {
        "iss": "https://gateway.example",
        "request_commit": "sha256:aaaaaaaaaaaaaaaaaaaaaaaaaaaaaaaaaaaaaaaaaaaaaaaaaaaaaaaaaaaaaaaa",
        "effective_request_commit": "sha256:eeeeeeeeeeeeeeeeeeeeeeeeeeeeeeeeeeeeeeeeeeeeeeeeeeeeeeeeeeeeeeee",
        "in_output_mode": "non_stream",
        "in_output_commit": "sha256:9999999999999999999999999999999999999999999999999999999999999999",
        "out_output_mode": "stream",
        "out_output_commit": "sha256:dddddddddddddddddddddddddddddddddddddddddddddddddddddddddddddddd",
        "policy": "repackage-as-stream/v1",
        "alg": "Ed25519",
        "kid": "gateway-key-9",
        "sig": "BASE64URL_SIGNATURE"
      }
    ],
    "output_commit": "sha256:dddddddddddddddddddddddddddddddddddddddddddddddddddddddddddddddd",
    "chunk_count": "42",
    "alg": "Ed25519",
    "kid": "gateway-key-9",
    "sig": "BASE64URL_SIGNATURE"
  }
}
\end{lstlisting}

\section{What \AEX\ Proves and Does Not Prove}
\begin{table}[h]
\centering
\caption{Claim boundary of \AEX.}
\label{tab:claims}
\begin{tabular}{p{0.43\textwidth}p{0.43\textwidth}}
\toprule
\textbf{AEX can prove} & \textbf{AEX does not prove} \\
\midrule
A trusted issuer signed a particular request-output relation. & The response is factually true. \\
A streamed output was received in the committed order and count, subject to the terminal proof mode that was actually used. & The provider used a specific internal model weight snapshot. \\
A request was transformed through a chain of explicitly signed, trusted transforms. & No hidden prompts, private retrieval, or server-side policy text existed. \\
A final complete output descended from a source issuer's complete output through an explicitly signed chain of trusted output transforms. & Every hidden intermediate output in that lineage is visible to the client. \\
A non-streaming or streaming response is inconsistent with the attested commitments if tampered with. & An accepted output transform is semantically correct, reasonable, or minimal absent external policy. \\
\bottomrule
\end{tabular}
\end{table}

\end{document}